\documentclass[journal=jacsat,manuscript=article]{achemso}

\usepackage[version=3]{mhchem} 
\usepackage{xcolor}
\usepackage{soul}

\usepackage{graphicx}
\usepackage{dcolumn}
\usepackage{bm}
\usepackage{float}
\usepackage{xcolor}
\usepackage{soul} 
\usepackage{siunitx}

\title{Using Limited Neural Networks to Assess Relative Mechanistic Influence on Shock Heating in Granular Solids}

\author{Brenden W. Hamilton}
\affiliation[Los Alamos National Laboratory]
{Theoretical Division, Los Alamos National Laboratory, Los Alamos, New Mexico 87545, USA}
\email{brenden@lanl.gov}
\author{Timothy C. Germann}

\date{\today}

\begin{document}

\begin{tocentry}
\begin{figure}[H]
  \includegraphics[width=0.6\textwidth]{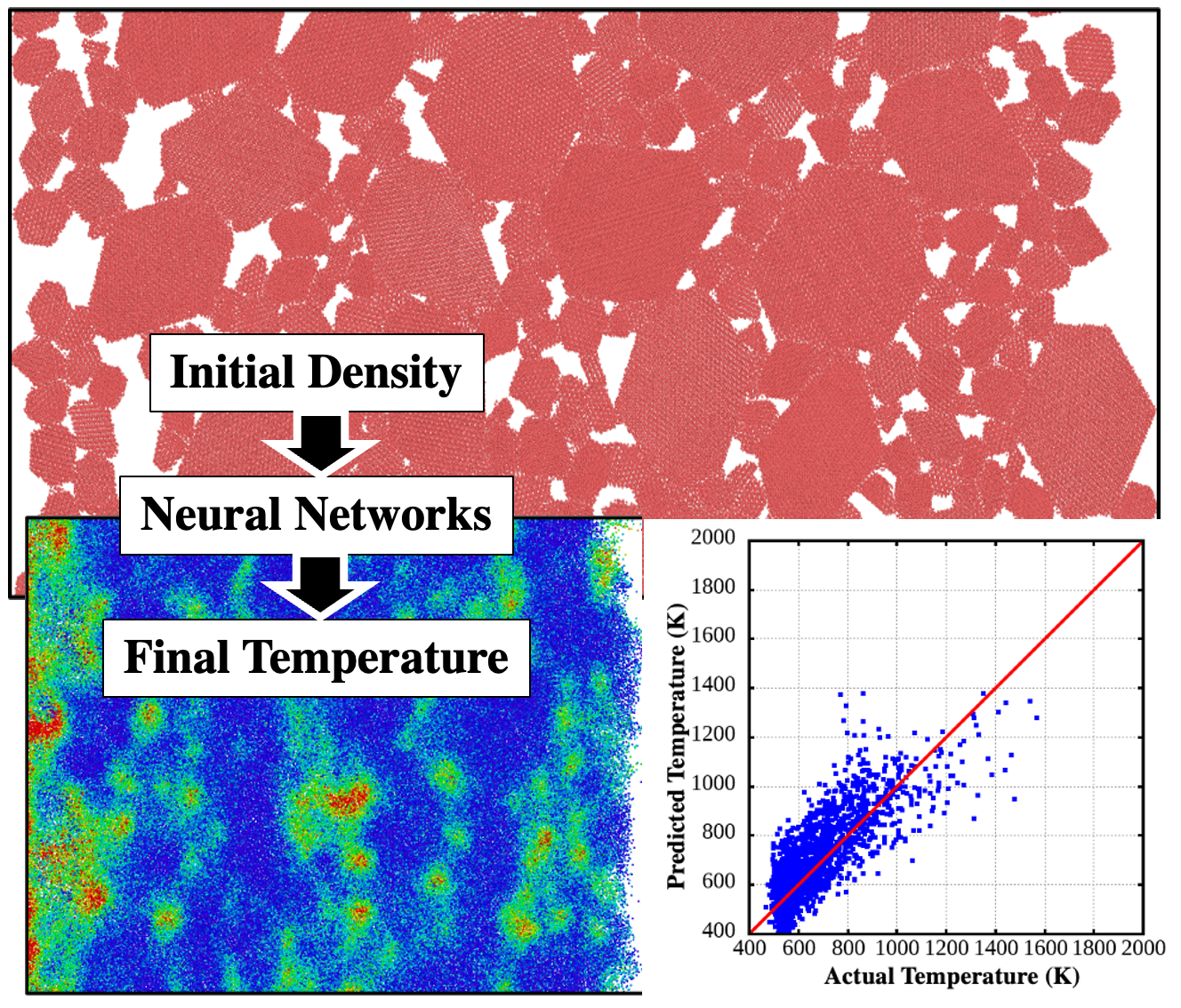}
  \label{fig:FgT}
\end{figure}
\end{tocentry}

\begin{abstract}

The rapid compaction of granular media results in localized heating that can induce chemical reactions, phase transformations, and melting.
However, there are numerous mechanisms in play that can be dependent on a variety of microstructural features.
Machine learning techniques such as neural networks offer a ubiquitous method to develop models for physical processes.
Limiting what kinds of microstructural information is used as input and assessing normalized changes in network error,
the relative importance of different mechanisms can be inferred.
Here we utilize binned, initial density information as network inputs to predict local shock heating in a granular high explosive trained
from large scale, molecular dynamics simulations.
The spatial extend of the density field used in the network is altered to assess the importance and relevant length scales of the physical mechanisms in play,
where different microstructural features result in different predictive capability.

\end{abstract}

\newpage
\begin{singlespace}

The rapid compaction of granular solids can lead to a wide variety of microstructural~\cite{Cherukara2014ShockLoading,brujic2005granular,lumay2005experimental,busignies2004compaction} 
and chemical~\cite{Cherukara2016ShockLoading,Strachan2003ShockWaves,yang1997shock,Zhou2019Shockwave} responses that are highly dictated by the initial local packing and structure of the material. 
While the use of molecular dynamics (MD) and continuum mechanics simulations have helped to elucidate the governing processes~\cite{Wood2018Multiscale,DavisHerring2010Effects,Rai2020Macroscale,Bidault2018Granularity,Xiong2019Molecular,pouliquen2003fluctuating,Kadau2007Shock,rutherford2017probing}, 
the wide range of mechanisms in play have prevented a unified understanding of events, especially the weighted relevance of various mechanisms.

A key example is the shock compression of energetic materials, in which the materials are typically utilized as neat or polymer bonded granular solids with a bimodal grain size distribution of larger grains and smaller ``fills"~\cite{Li2022PBX,Skidmore1998Evolution,Willey2006Changes}.
The shock initiation of chemistry, which can lead to a run to detonation, is governed by the formation of localizations of excess energy known as hotspots, which are typically defined by their temperature and size~\cite{Hamilton2021Review,Handley2018Review}.
These hotspots form through shock induced processes such as intra- and inter-granular void collapse, shear band formation, jetting of material, and inter-granular friction~\cite{Davis1981HotspotsLANL,Li2020HotspotCracks,Cawkwell2008ShearBand,Kroonblawd2020ShearBands,Dienes1985StatisticalCrack}.
From system to system, these individual processes can be influenced by material orientations, crystal defect formation, surface properties, and void shapes and sizes~\cite{Zhao2020Tandem,Hamilton2022PEHotspot,Lafourcade2018Irreversible,Grilli2018EffectOrientation,li2022shock,Bidault2018Impact,hamilton2023energy}.
Void collapse is typically the dominant process, with the energy localization increasing with increasing P-V work done~\cite{Holian2002AtomisticModelsHotspot}.
Hence, broadly understanding the overall shock compression and initiation involves a wide range of materials models and highly detailed structural information.

Additionally, shock compaction not only localizes thermal energy, but can also deform individual molecules, causing them to exist in strained states~\cite{Hamilton2021HotspotsBetterHalf}. 
These intra-molecular strains can alter reaction kinetics and pathways through mechanochemistry~\cite{Hamilton2022Extemporaneous,Hamilton2022ManyBody,hamilton2023interplay}.
Interestingly, these strain energies are thought to occur through fundamentally different processes than temperature localization~\cite{hamilton2023energy,hamilton2023intergranular}.
Being able to predict the localization of both temperature and strain energy in a hotspot remains a grand challenge for the energetic materials community and is highly relevant to general materials compaction problems.
Being able to predict these processes without running computationally expensive molecular dynamics and hydro-code simulations, as well as to better evaluate the key/necessary mechanisms, is crucial to the materials physics
and condensed matter chemistry communities.

Materials science has recently experienced a rapid increase in the use of machine learning (ML) to extract and understand physical processes that can occur~\cite{Iten2020Discovering,butler2018machine,zhong2022explainable}. 
ML has played a key role in the development of computational models~\cite{huan2017universal,Chan2019Machine,zuo2020performance,Yoo2021NNRF,Lindsey2017Chimes,hamilton2023high,zhang2019embedded}, predicting properties,~\cite{fernandez2014rapid,xie2018crystal,Sakano2020Unsupervised,ward2016general,zhuo2018predicting} 
and characterizing materials~\cite{carbone2020machine,timoshenko2017supervised,yao2022autophasenn,hu2019machine}. While predictions from large and non-linear neural networks typically function as a black box, limiting and altering the 
physical information that informs the network can help to tease out which properties and mechanisms are critical to a processes by the network's ability to 
make predictions given its limited subset of input information. This process, employed here, is similar to a "leave one feature out" scheme.

Here, we utilize non-reactive, all-atom MD simulations to model the shock response of the granular high explosive 1,3,5-triamino-2,4,6-trinitrobenzene (TATB). Simulation details are provided in the Methods section and  Supplemental Materials section SM-1.
A neural network is used to predict the final temperature and intra-molecular strain energy fields
given just the initial density field of the unshocked system. The level of coarsening and total extent of the density information given is varied to assess the amount of local information needed to properly predict energy localization 
from granular compaction. Increasing density resolution provides more information pertaining to pore shape and local curvature, but using only density restricts
potentially critical information such as local orientation and crystalline defects.
It should be noted that the purpose of this work is not to minimize the error of the networks and make the best model possible. It is to systematically change
the inputs given to the network such that physical trends and important physico-chemical mechanisms can be inferred from the relative change in each
network's predictive ability. However, this methodology does necessitate the network predictions be reasonable accurate as a baseline for extracting materials physics, such that the model is presumable learning some description of the physics.

Predictions of temperature ($T$) and intra-molecular strain energy ($U_{Latent}$) fields are done with a simple, non-linear neural network.
The system is binned in a Lagrangian fashion on the initial frame prior to shock to provide the density descriptors. From the simulation, each molecule has a defined time is is "shocked", and a "composite frame" is constructed by taking 
the position and thermodynamics at each molecule at 5ps after its shocked time, i.e. molecules are taken from different simulation frames such that they are at the same relative time compared to being compressed.

The input layer consists of the density of a bin and its $N$ sets of nearest neighbors (initial frame), as shown in Figure 1 for the $N=1$ case (a second shell of bins would be $N=2$ and just the central bin, no neighbors, is $N=0$).
Bins with zero molecules are considered as neighbor bins but not as center bins as they will not have a final temperature/energy for an output.
The output layer is either the $T$ or $U_{Latent}$ of the center bin (composite frame); each of these two values is trained with separate networks of the same architecture. 
All networks consist of an input layer of size $(2N+1)^2$, an output layer of size 1, and a single hidden layer of size $\lfloor0.5(2N+1)^2\rfloor$.  
Figure 1 exemplifies how the all-atom structure is encoded into the neural network input layer, where the output layer corresponds to a region of the all-atom results, the same center Lagrangian bin, which has mean $T$ and $U_{Latent}$ values.
Additional details on network architecture are available in the Methods section and Supplemental Materials section SM-2.

\begin{figure}[htbp]
  \includegraphics[width=0.5\textwidth]{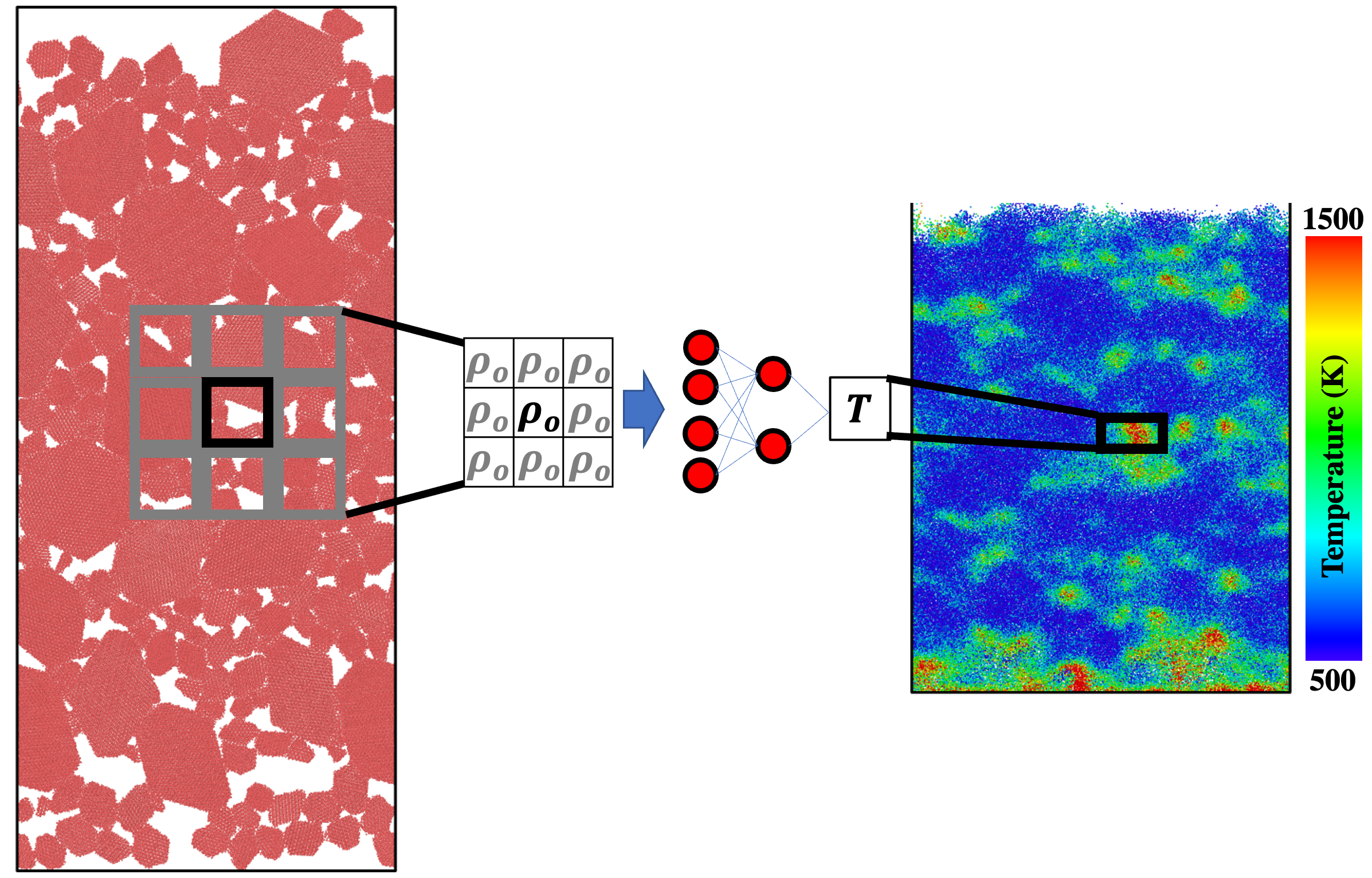}
  \caption{Initial configuration and composite temperature map for the testing set system with example binning for network input and output.}
  \label{fig:Fg1}
\end{figure}

Different square bin sizes of 2.5 nm, 3.0 nm, 4.0 nm and 5.0 nm are used. To compare different bin size results, we define the spatial extent $(SE)$ of the input layer, where $SE = L(N+\frac{1}{2})$ where 
$L$ is the bin length and $N$ is the number of sets of nearest neighbor bins included. This is the equivalent to the radius of an inscribed circle for the total square of bins used.
Figure 2 shows parity plots of $T$ and $U_{Latent}$
for $L$ = 5.0 nm and 2.5 nm with nearly equivalent spatial extents of 27.5 and 26.25 nm, respectively.
For the temperature (left column plots), both show decent correlation with the parity line, with RMS errors of 84.4 K and 108.9 K for 5.0 and 2.5 nm bins, respectively, where peak (individual) molecular temperatures in the hotspots are near 
2000 K and 600-700 K in the bulk shocked materials.

When comparing the two sets at different bin sizes, it is crucial to consider that while the 2.5 nm bins results in four times as many training and testing points, it also samples a much wider
range of values, as the smaller bins provide less smoothing of extreme temperatures. The 2.5 nm cases have considerably more values above 1000 K and its peak hotspot temperatures are much higher than the bulk temperature, relative to the 5.0 nm bins. 
Both cases mostly have slight over-predictions of values around 800 K and under-predictions for those above 1000 K. 

\begin{figure}[htbp]
  \includegraphics[width=0.5\textwidth]{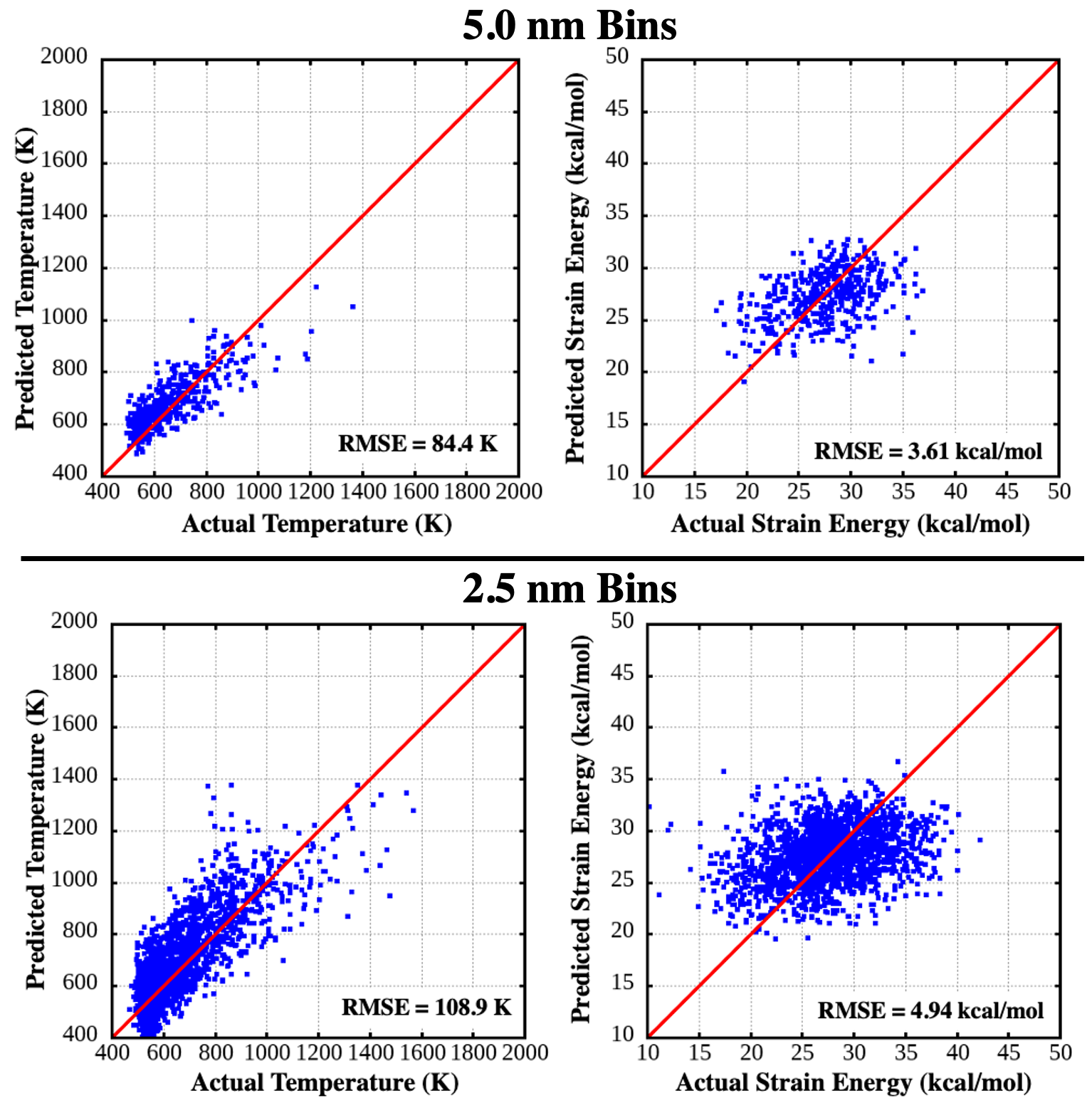}
  \caption{Parity plots of $T$ and $U_{Latent}$ for the 5.0 nm bins and 2.5 nm bins for spatial extents of 27.5 and 26.25 nm, respectively. Data consists of predictions of the smaller MD cell used exclusively for testing.}
  \label{fig:Fg2}
\end{figure}

For the $U_{Latent}$ predictions shown in Figure 2, there is considerably less correlation with the parity line. The predictions even appear to be less correlated overall at the smaller bin sizes. While the $U_{Latent}$ fields, shown in Supplemental Materials section SM-4, are more disperse that the $T$ fields,
there are still notable regions of higher and lower strain energy. The predicted fields are considerably more homogeneous and noisy, showing that the density field alone is not enough to predict the mechanisms that drive molecules to bent and distorted shapes, and that more complicated microstructural or thermodynamic information
is likely needed to make these predictions. This additionally helps to verify previous results that concluded that the $T$ and $U_{Latent}$ forming mechanisms are different, as their localizations do not occur on a one-to-one basis~\cite{Hamilton2021HotspotsBetterHalf}.
$U_{Latent}$ also has a strong influence from pore size at larger pores~\cite{hamilton2023energy}, and the system sizes here may not be large enough to provide a wide enough range of examples in the training set. Compared to temperature, the density (and therefore P-V work) alone 
is not enough to make decent quantitative predictions of the $U_{Latent}$ field.

Figure 3 shows heat maps of the actual and predicted temperatures, as well as the difference,
for a 2.5 nm bin case. Other bin size and nearest neighbor cases are available in Supplemental Materials section SM-3. From this, it can be assessed that hotspots that are longer in the shock direction, like points (a) and (b) in Figure 3, are under-predicted. However, hotspots that are longer in
the cross direction, like points (c) and (d) in Figure 3, are over-predicted. The initial microstructure of this case is shown in Figure 1. Vertical pores and high aspect curvature pores can often result in molecular jetting, leading to high levels of expansion before getting re-compressed~\cite{germann2002DetSymp,Li2020HotspotCracks,li2022shock},
however, wider or more circular pores often result in a more hydrodynamic or plastic flow type responses that incurs much less P-V work during re-compression relative to the jetted material, especially as material will have less physical space to expand into the pore before getting re-compressed. It should be noted that, while there is under predictions of the hottest hotspots and over prediction of the colder ones, the model still generally predicts vertical pores to be hotter than the horizontal ones.

\begin{figure}[htbp]
  \includegraphics[width=0.95\textwidth]{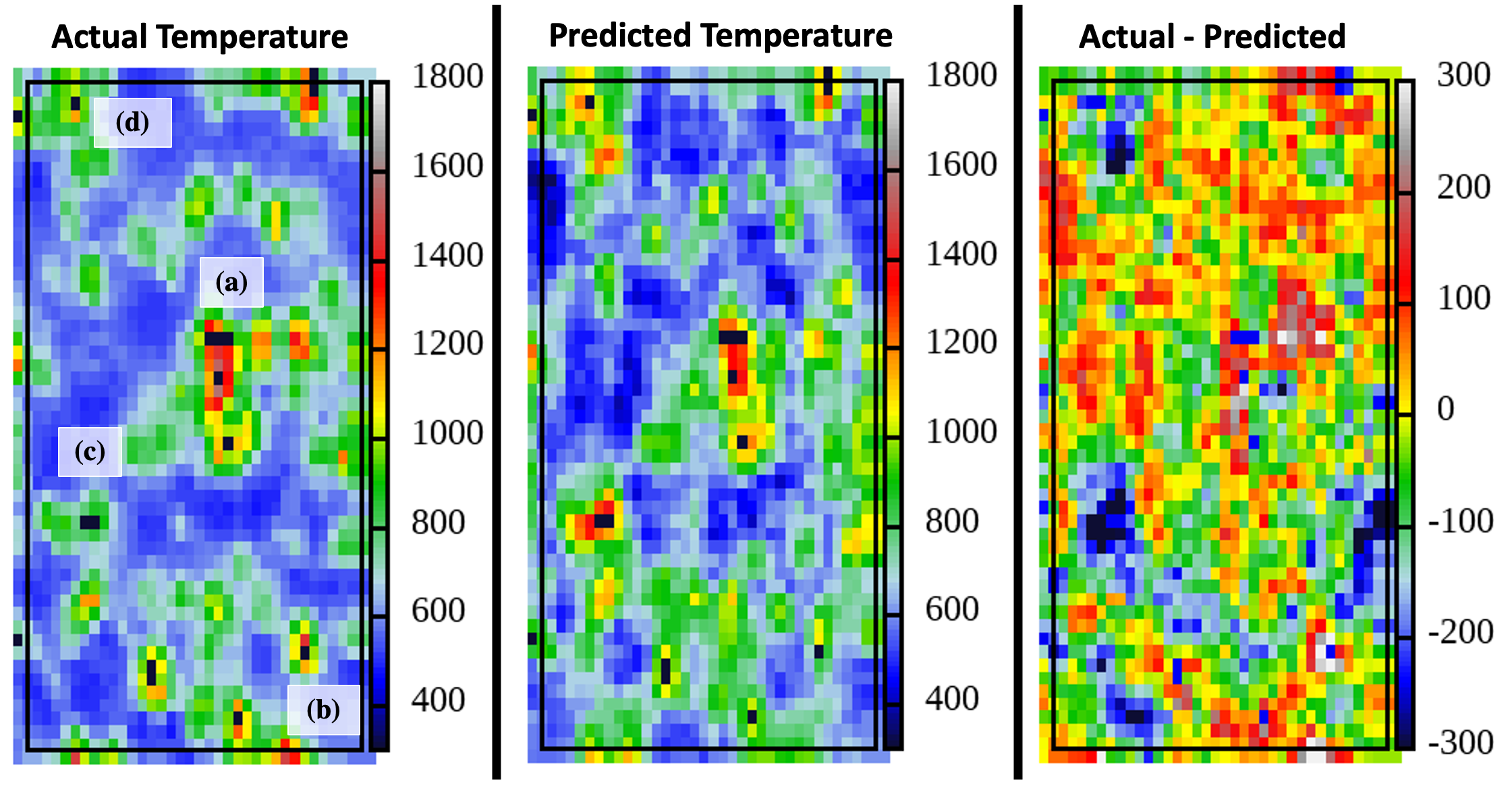}
  \caption{Heat maps of the actual and predicted temperatures, as well as the difference between the two, for the 2.5 nm bin case with a 26.25 nm spatial extent. Points a-d on the actual temperature map correspond to specific hotspots where a and b are under-predicted, and c and d are over-predicted.}
  \label{fig:Fg3}
\end{figure}

The two key factors in qualitatively predicting hotspot temperatures from pore collapse are the pore size and shape~\cite{Wood2018Multiscale,Li2020HotspotCracks,hamilton2023energy}.
Based on these results, the predictions here appear to be fully considering size with the largest pores (points (a) and (c)) predicting high temperatures and the smaller pores typically predicting lower temperatures. 
Molecular jetting, which greatly influences material expansion into pores, is a much more complex mechanical process where things like curvature of the pore 
come into play~\cite{Li2020HotspotCracks,li2022shock}. Even with small initial bins, this information is partially coarsened out, preventing the network from making these predictions. Information on the shock wave structure/shape, which will change as it progresses over the material, is also unknown to the network and can cause different amounts of shock focusing that leads to jetting.
Additionally, the network does not have information related to crystal orientation, which results in changes
in hotspot temperature on the order of the errors shown here~\cite{Hamilton2022PEHotspot}, however, the grain orientations chosen here minimize anisotropy between grains. 
Hence, while P-V work from pore size is enough to make decently quantitative predictions of hotspot temperature, the finer
microstructural details are likely needed to correct for errors on the order of several hundred Kelvin.
It is most likely a combination of these omitted features that drives the error in predictions, especially the under-prediction of the highest temperature hotspots.

Figure 4 shows root mean square (RMS) errors and Linear Normalized RMS errors (LN-RMSE) for all networks trained. The LN-RMSE are RMSE values normalized by the RMSE of the linear regression between the density of a bin
(no nearest neighbor information) and the temperature. For the top row in Figure 4, which is temperature, we interestingly see, for small to intermediate spatial extent, similar RMSE from all bin sizes, followed by a divergence of values.
The slight uptick in errors for large spatial extent are attributed to a static stopping criteria based on error reduction over the previous 200 epochs.
For larger SE, the network itself is proportionally larger and potentially learns at a slower rate, especially for smaller bin sizes where $N$ is much larger for a given SE and the network size grows as $N^2$.

For an equal spatial extent between two different bin sizes, the smaller bin will result in more training/testing data for a given simulation size. However, the smaller bin system is also less coarse grained, and will have a
wider range of peak temperatures and more pronounced temperature gradients. Hence, with more training data and more resolved microstructural data from the density input, the smaller bins are able to reach roughly the same
level of predictive power as for the much smoother fields of the larger bins. For the LN-RMSE, as smaller bins will have more variability in temperature with density and therefore a larger normalizing constant, the
smallest bin cases perform the best. This normalized case shows the predictive power of adding more microstructural information with a finer density field. As there is significantly less accuracy in the $U_{Latent}$ predictions,
the trends of decreasing errors with SE and various bin size effects are less physically meaningful. Plots of RMSE and LN-RMSE for $U_{Latent}$ are available in Supplemental Materials section SM-5. These show similar trends with SE
to temperature, yet opposite trends with respect to bin size, which is most likely an effect of coarsening significant noise and the ease of predicting a more uniform field.

\begin{figure}[htbp]
  \includegraphics[width=0.95\textwidth]{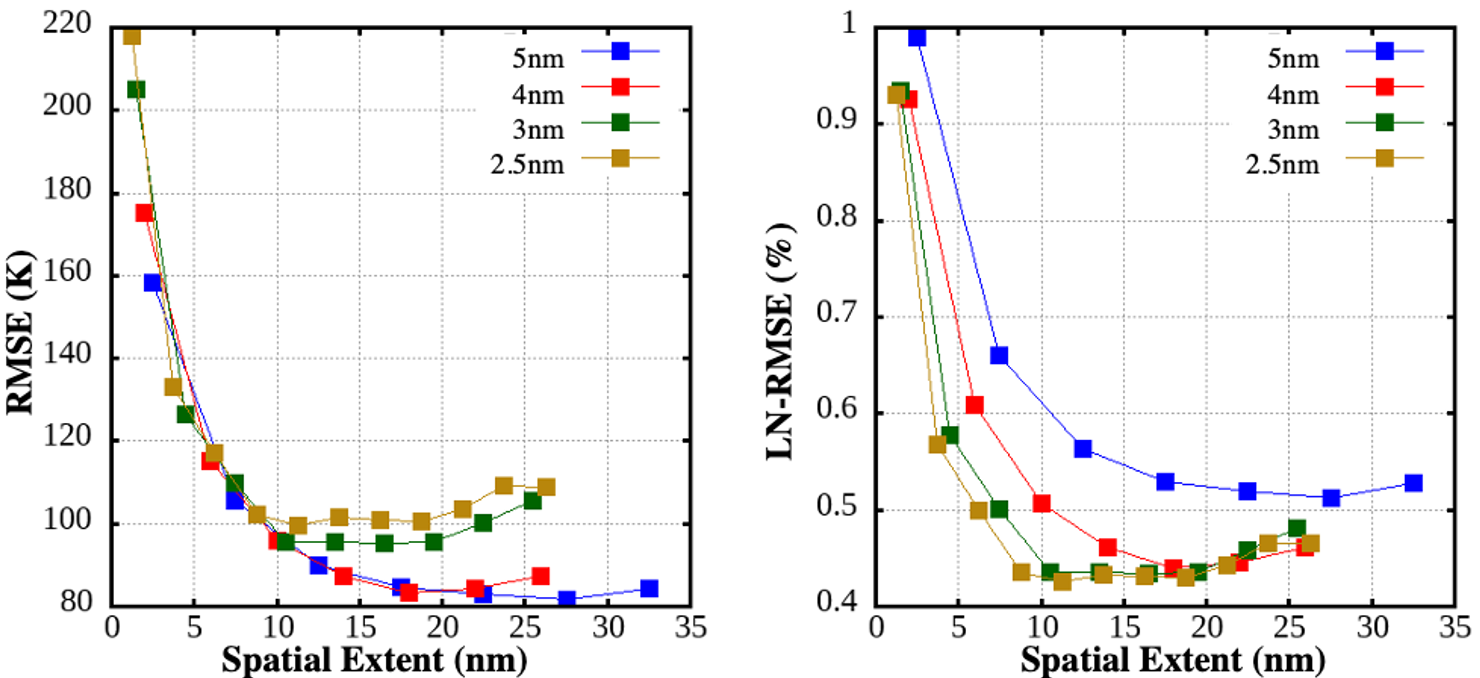}
  \caption{Root mean square errors for all temperature networks run, as a function of spatial extent. Colored lines represent different bin sizes. LN-RMSE (Linear Normalized RMSE) are
  RMS errors that are normalized by the RMS error of a linear network for zero nearest neighbors ($N=0$).}
  \label{fig:Fg4}
\end{figure}

In summary, MD simulations of shock compaction of a granular material resulted in heterogeneous localization of both temperature and intra-molecular strain energy, where the latter is known to be responsible for mechanochemical effects.
A Lagrangian binning of initial microstructures was used to embed local density information, but intentionally excludes information pertaining to local crystal
structure, defects, orientation, and incident wave structure.
These density bins were utilized as an input layer to a neural network to predict the shock induced $T$ and $U_{Latent}$ fields.

From trends in RMS errors for different bin sizes and spatial extents of the input layer, we find that the predictability of a network increases with more spatial extent, as well as with smaller bins which would carry more precise 
microstructural information. These improvements are despite the smaller bins leading to much larger variability and fluctuation in the temperature fields, where larger bins smooth and coarsen that information.

While density information did not allow for accurate prediction of $U_{Latent}$, the networks readily predicted the temperature fields with some of the hotspots being over-predicted, while others are under-predicted. 
While the network
predicts hotter temperatures for larger pores, where more P-V work can occur during compaction, it fails to fully capture other mechanisms such as jetting and molecular ejecta, which can lead to extreme temperatures. This failure manifests in pores
that nucleate these jetting mechanisms, those elongated in the shock direction, to have under-predicted temperatures. Additionally, equi-axed or wider pores that do not jet are slightly over-predicted. This leads the network to give less variation in prediction
from hotspot to hotspot, but still captrues the general trend with pore size. 

Hence, we are able to utilize neural network predictions to show that, while P-V work is the dominant mechanism in temperature
localization, not having finer microstructural details such as those on the nanometer to sub-nanometer length scale leads to errors with
the hotspot on the order of a few hundred Kelvin. Additionally, P-V work and pore shape are shown to be much less important mechanisms for 
the $U_{Latent}$ field, which may have considerable influence from plasticity levels and material flow rate~\cite{Kroonblawd2020ShearBands,hamilton2023energy}.

This work shows promise in the use of these limited neural networks to assess physical mechanisms in play for complex, physico-chemical processes in
condensed matter systems. In future work, a wider variety of input descriptors can be utilize that, in addition to density,
map features such as crystal orientation, pre-existing crystalline defects, surface roughness in pores,
and the structure/shape of the incident shock wave which will be altered from upstream microstructural features and shock instabilities.
By coupling these features with a 'leave one feature out' type scheme and the varying spatial extend scheme used here, a relative importance of key hotspot mechanisms and their necessary descriptors can be deduced.

\section{Methods}

All simulations were run with all-atom MD using the LAMMPS software package~\cite{Plimpton1995LAMMPS,Thompson2022LAMMPS}.
Interatomic interactions were calculated using the nonreactive, nonpolarizable forcefield from Bedrov et al.~\cite{Bedrov2009Molecular}, with tailored
harmonic bond stretch and angle bend terms~\cite{Kroonblawd2013Theoretical} and an intramolecular O-H repulsion term~\cite{Mathew2015AnisotropySurface}.
Electrostatics were solved for in real space with the Wolf potential~\cite{Wolf1999Exact}.
Van der Waals interactions were modeled using the Buckingham potential. All simulations are conducted with a 0.25 fs timestep.

Simulations cells of granular TATB were built with the PBXGen algorithm~\cite{Li2022PBX} using columnar grains with a bimodal grain size distribution with peaks at 40 nm and 8 nm. 
The smaller 'fill' grains make up roughly 2/3 of the total grain count.
All grains are crystallographically oriented so that the TATB [001] direction is in the periodic $Z$ direction with thickness 4.1 nm (into the page in the Figures 1 and 3), which minimizes grain-to-grain anisotropy.
Two cells were constructed with different sizes of 100 x 400 nm (12440592 atoms),
and 100 x 200 nm (6183984 atoms),
where the larger system is used to make up the training set and smaller system is the testing set

Shock simulations are conducted using the reverse ballistic method with a particle velocity of 2.0 km/s, with the resulting shock traveling in the $+Y$ direction (upwards in the Figures 1 and 3).
The $Y$ direction is a free boundary, whereas the other directions are periodic.
Analysis is conducted using a per molecule basis, using the molecular center of mass as its position.
Temperature ($T$) is the molecular roto-vibrational kinetic energy in units of Kelvin. 
Intra-molecular strain energy ($U_{Latent}$) is defined as the excess intra-molecular potential energy with respect to the equi-partition theorem~\cite{Hamilton2022PEHotspot}.
Supplemental Materials section SM-1 provides additional MD and PBXGen methods.

For the neural networks, a sigmoid function is used into the hidden layer, and a linear function into the output layer. A bias is allowed for both layers. Supplemental Materials section SM-2 provides additional machine learning method details.

\section{Acknowledgments}
Funding for this project was provided by the Director’s Postdoctoral Fellowship program at Los Alamos National Laboratory, project LDRD 20220705PRD1. Partial funding was provided by the Advanced Simulation and Computing Physics and Engineering Models project (ASC-PEM). This research used resources provided by the Los Alamos National Laboratory (LANL) Institutional Computing Program. This work was supported by the U.S. Department of Energy (DOE) through LANL, which is operated by Triad National Security, LLC, for the National Nuclear Security Administration of the U.S. Department of Energy (Contract No. 89233218CNA000001). Approved for Unlimited Release LA-UR-23-24466.

\section{Supplemental Material}
\includegraphics[scale=0.7,page=2]{Supplemental_Material.pdf}
\newpage
\includegraphics[scale=0.7,page=3]{Supplemental_Material.pdf}
\newpage
\includegraphics[scale=0.7,page=4]{Supplemental_Material.pdf}
\newpage
\includegraphics[scale=0.7,page=5]{Supplemental_Material.pdf}
\newpage
\includegraphics[scale=0.7,page=6]{Supplemental_Material.pdf}
\newpage
\includegraphics[scale=0.7,page=7]{Supplemental_Material.pdf}
\newpage
\includegraphics[scale=0.7,page=8]{Supplemental_Material.pdf}
\newpage
\includegraphics[scale=0.7,page=9]{Supplemental_Material.pdf}
\newpage

\bibliography{references}
\end{singlespace}
\end{document}